\documentclass[aps,prl,twocolumn,preprintnumbers,amsmath,superscriptaddress]{revtex4}
\usepackage{bm}
\usepackage{amsmath}
\usepackage[dvips]{graphicx}
\usepackage{color}

\begin{document}
\bibliographystyle{apsrev}
\title{ AC Hopping Magnetotransport Across the 
Spin Flop Transition in Lightly Doped La$_{2}$CuO$_4$}

\author{Oleg P. Sushkov}
\affiliation{School of Physics, University of New South Wales, Sydney 2052, Australia}
\author{Valeri N. Kotov}
\affiliation{Department of Physics, Boston University, 590 Commonwealth Avenue, Boston, MA 02215}

\begin{abstract}
The weak ferromagnetism present in insulating La$_{2}$CuO$_4$ at low doping leads
to a spin flop transition, and to  transverse (interplane)  hopping of holes in a  strong
external magnetic field. This results in a dimensional crossover 2D $\to$  3D 
for the in-plane transport, which in turn leads to  an increase of the 
hole's localization length and increased conduction. We demonstrate 
theoretically that as a consequence of this mechanism, a frequency-dependent 
jump of the in-plane ac hopping  conductivity occurs at  the spin flop 
transition.  We predict the value and the frequency dependence of the jump.
Experimental studies  of this effect would provide  important confirmation
of the emerging understanding of lightly doped insulating La$_{2-x}$Sr$_x$CuO$_4$.

\end{abstract}
\maketitle

{\it{Introduction.}}
The interplay between the charge and spin degrees of freedom is crucial for 
our understanding  of the high-temperature superconductivity.
 A lot can be learned about this interplay 
 by studying the insulating phases of the copper oxides.
Lightly doped La$_{2}$CuO$_4$ is especially interesting because of 
the  weak ferromagnetism which exists in the N\'eel phase.
The weak ferromagnetism provides a ``handle'' that 
allows one to influence the spins by application of a moderate magnetic field
 and hence to study how the charge dynamics is influenced by the spins. 
This issue has been recently addressed both experimentally
\cite{Thio,Chen,Ando,Ono,Gozar,Keimer} and theoretically \cite{CP,Lara,LS}.

The present work is relevant to La$_{2-x}$Sr$_x$CuO$_4$,
La$_{2}$Cu$_{1-x}$Li$_x$O$_4$, and La$_{2}$CuO$_{4+y}$ at very low doping.
Generally these compounds have very different properties.
La$_{2-x}$Sr$_x$CuO$_4$ has an insulator-superconductor transition 
at $x \approx 0.055$ \cite{Kastner}  while La$_{2}$Cu$_{1-x}$Li$_x$O$_4$ 
remains an insulator at all dopings \cite{Sarrao}.
Elastic and inelastic neutron scattering in La$_{2-x}$Sr$_x$CuO$_4$ 
reveals incommensurate magnetic peaks
\cite{Yamada,Wakimoto,Matsuda,Fujita} while neutron scattering in
La$_{2}$Cu$_{1-x}$Li$_x$O$_4$ demonstrates only an inelastic peak that is
always commensurate \cite{Bao,Chen1,Chen2}.
There are, however, similarities between these compounds: the long-range
N\'eel order is destroyed at rather close values of doping, $x =0.02$ in
La$_{2-x}$Sr$_x$CuO$_4$,  and $x =0.03$ in La$_{2}$Cu$_{1-x}$Li$_x$O$_4$ 
\cite{Kastner,Heffner}. The most important similarity is that at low doping
 all compounds exhibit  variable-range hopping (VRH) 
conductivity \cite{Chen,Kastner} that unambiguously indicates localization of holes.
Localized holes in La$_{2-x}$Sr$_x$CuO$_4$  can lead to a diagonal spiral distortion of the
spin background, and the corresponding theory \cite{HCMS,juricic04,SK,JSNMS,LMMS,luscher} explains 
quantitatively the whole variety of magnetic data.
The transport theory based on localization \cite{KS,kotov06a}
explains quantitatively the in-plane anisotropy
of the dc and ac conductivities in La$_{2-x}$Sr$_x$CuO$_4$
as well as the negative  in-plane dc magnetoresistance in La$_{2-x}$Sr$_x$CuO$_4$
 and La$_{2}$CuO$_{4+y}$.

 In the low-doping region there exist anisotropies in the spin-spin
interactions, such as Dzyaloshinsky-Moriya (DM) and XY terms,
 in addition to the Heisenberg exchange.  In the
N\'eel phase the anisotropies confine the spins to the $(ab)$ plane and fix 
the direction of the N\'eel vector to the $\hat{b}$-orthorhombic axis. 
Moreover, the DM interaction induces a small out-of-plane spin component that is
ferromagnetic in the plane (weak ferromagnetism) but staggered in the
out-of-plane $\hat{c}$-direction. This component can be easily influenced
by an external magnetic field applied in different directions.

Magnetic field directed along the $\hat{c}$ axis can cause an alignment
of the out-of-plane moments via a spin flop transition at a critical field
$H_{f}$, determined by the competition between the DM and inter-layer
Heisenberg exchange.  Typically $H_{f} \approx 5-7 \ {\mbox T}$
\cite{Ando,Gozar,Keimer}.
Theoretically a more complicated behavior than just a spin flop is 
also possible \cite{CP}. Which particular regime is realized
depends on the values of the anisotropy parameters.
The experimental data are not 100\% concusive, but still they mostly indicate
a simple spin flop and this is the picture that we accept in the present work.
Magnetic field directed along the orthorhombic $\hat{b}$ axis causes a continuous
rotation of the spins towards the $\hat{c}$ axis~\cite{Thio,Ono,Gozar,Lara,LS,ML}.
The spins align completely along $\hat{c}$ at a field $H_{c2}\sim 20 \ {\mbox T}$. 
An intermediate in-plane spin flop is also possible at $H_{c1}<H_{c2}$ 
\cite{Lara,LS}. The intermediate in-plane spin flop is very sensitive to
doping and the situation is different for Sr, Li,  and O doping~\cite{ML}.
However the intermediate in-plane spin flop practically does
not influence the transport properties.

In the present work we study the  in-plane ac magnetoresistance (MR) 
across the spin flop transition, and our results are applicable to 
La$_{2-x}$Sr$_x$CuO$_4$, La$_{2}$Cu$_{1-x}$Li$_x$O$_4$, and La$_{2}$CuO$_{4+y}$.
Doping is assumed to be so small that the compounds are in the N\'eel phase
(the weak ferromagnetism regime). For definitiveness and simplicity  we concentrate on the
case of a field along the  $\hat{c}$ direction where the spin flop transition is sharp,
 although an extension for a field in the  $\hat{b}$ direction is straightforward.
 We rely on a recently developed theory that provides a detailed knowledge of the
 evolution of the hole's localized wave-function across the spin flop transition  
\cite{kotov06a}. This evolution takes into account the opening up of an 
 interplane hopping channel at the spin flop and was used to explain the
 negative dc MR, in excellent agreement with experiment \cite{kotov06a}.
 Here we calculate the ac MR within this theoretical framework, and show
 that a large negative MR is also expected in this case. 
 The MR value depends on the mechanism of ac hopping conduction, 
 determined by the value of the ac frequency $\omega$ compared
to the temperature $T$.
 The variation of the ac MR with $\omega$  is found to be slow,
 due to the logarithmic dependence of the VRH  distance
 on $\omega$. Moreover, at low frequency a saturation of the
 MR is predicted. These features distinguish the ac regime from the
 dc case which has been studied more
 extensively in experiment. 

{\it{Jump in the in-plane ac conductivity at the spin-flop transition for 
field $H \parallel \hat{c}$. Zero temperature phononless case.}}
First we  consider the quantum, phononless regime that corresponds to very low
 temperatures ($T\ll\omega$).
In this case the ac absorption is due to  resonant electromagnetic transitions between
hybridized symmetric and antisymmetric states formed by a  hole bound 
to two impurities separated by distance $r$.
It was demonstrated in Ref.~\cite{kotov06a} that at
$H < H_f$  the hole hopping matrix element between the CuO$_2$ planes
is extremely small, and therefore we have a pure two-dimensional (2D) situation.
According to Shklovskii  and Efros \cite{SE}, the conductivity
in the quantum regime (which also takes into account the Coulomb
 interaction $ e^2/r$ within the resonant pair),
appropriately modified for our 2D case, is
\begin{equation}
\label{se}
\sigma (\omega) \propto \int_{r_{\omega}}^\infty
\left(\omega+\frac{e^2}{r}\right)
\frac{r^3I(r)dr}{\sqrt{\omega^2-4I(r)}}  \ ,
\end{equation}
where 
\begin{equation}
\label{tp0}
I(r)=I_0^2P(r) \ . 
\end{equation}
Here $I(r)$ is the overlap integral of the two
 localized states, and  $I_0\sim \epsilon_0 \approx 10 {\mbox{meV}}$, where $\epsilon_0$ is the hole
binding energy~\cite{kotov06a}. The tunneling probability $P(r)$ is
\begin{equation}
\label{tp}
P(r)=e^{-2\kappa r} \ ,
\end{equation}
where $\kappa$ is the inverse localization length. From
 now on we  omit the exact prefactors in formulas like (\ref{se})
since we will be  interested only in conductivity ratios.
The lower limit of integration $r_{\omega}$ in (\ref{se})
is determined by the zero of the  denominator of the integrand.
Performing the integration in (\ref{se}) with logarithmic accuracy,
one obtains  the well known answer \cite{SE}
\begin{equation}
\label{ES}
\sigma(\omega)\propto \frac{\omega}{\kappa}\left(\omega+\frac{e^2}{r_{\omega}}\right)
r_{\omega}^3 \ ,
\end{equation}
where
\begin{equation}
\label{roo}
r_{\omega}=\frac{1}{\kappa}\ln\left(\frac{2\epsilon_0}{\omega}\right).
\end{equation}

 Now we turn to the case  $H > H_f$.
 As discussed in Ref.~\cite{kotov06a},  the spin flop  gives rise to hopping
in the $\hat{c}$-direction and this effectively changes the dimensionality of the bound state, 
increasing the localization length. The hopping in the $\hat{c}$-direction is described
by the parameter $Zt_{\perp}\sim 0.5 \mbox{meV}$, where $t_{\perp}$ is the effective hopping
matrix element, and $Z\approx 0.3$ is the hole quasiparticle residue determined by 
spin quantum fluctuations. We estimate the hopping relative to the binding energy 
\begin{equation}
\label{zte}
 \frac{2Z t_{\perp}}{\epsilon_0} \sim 0.1 - 0.2 \ .
\end{equation}
Due to the dimensional crossover the wave function of the hole bound state is changed in a
peculiar way. The general integral representation for the probability to find
the hole at a distance $r$ from the trapping center is derived in \cite{kotov06a}. 
The result  is simplified greatly in two  limiting regimes:
(1) Very large distances, $\kappa r \gg \epsilon_0/(2Zt_{\perp})$, and 
(2) Intermediate large distances, $\epsilon_0/(2Zt_{\perp}) \gg\kappa r \gg 1$.
The typical hopping distance is given by (\ref{roo}), and  therefore the 
Very large distance regime is realized  for 
$\ln\left(\frac{2\epsilon_0}{\omega}\right) \gg \epsilon_0/(2Zt_{\perp})$, and  the
Intermediate large distance regime is  valid for 
$\epsilon_0/(2Zt_{\perp}) \gg \ln\left(\frac{2\epsilon_0}{\omega}\right) \gg 1$.

In the Very large distance regime,
the main result of Ref.~\cite{kotov06a} is that the localization
length increases to
\begin{equation}
 \tilde{\kappa}^{-1} = \kappa^{-1} [1-(4Zt_{\perp}/\epsilon_0)]^{-1/2},\ H>H_{f}.
\end{equation}
 The tunneling probability is basically given by
the same Eq.~(\ref{tp}) with the replacement $\kappa \to  \tilde{\kappa}$. 
Therefore Eq.~(\ref{ES}) is also valid and hence
the ratio of conductivities after and before the flop is
\begin{eqnarray}
\label{vld}
\frac{\sigma_{H > H_f}}{\sigma_{H < H_f}}&=&\left(1-\frac{4Zt_{\perp}}{\epsilon_0}\right)^{-2} 
\left [ \frac{1+ V_{\omega}  \sqrt{1-\frac{4Zt_{\perp}}{\epsilon_0}}}{1+V_{\omega} } \right ]  , \\
&& \kappa r_{\omega}  \gg  \epsilon_0/(2Zt_{\perp}). \nonumber
\end{eqnarray}
We have introduced here the relative strength of the Coulomb interaction
\begin{equation}
V_{\omega} \equiv \frac{e^{2}/r_{\omega}}{\omega} = 
\frac{\epsilon_{0}}{\omega \ln{(2\epsilon_0/\omega)}} \ ,
\end{equation}
and taken into account that $\epsilon_{0} = e^{2}\kappa$.
 For typical frequencies the Coulomb interaction  always dominates, $V_{\omega}\gg 1$.
 It is clear that the conductivity increases after the flop, $\sigma_{H > H_f}/\sigma_{H < H_f} >1$,
 i.e. the magnetoresistance is negative.

In the Intermediate large distance regime the propagation probability
is \cite{kotov06a}
\begin{equation}
\label{pid}
P(r) = e^{-2\kappa r}\left(1+4(\kappa r)^2\frac{(Zt_{\perp})^2}{\epsilon_0^2}\right)\ .
\end{equation}
One has to substitute this expression into  Eq.~(\ref{se})
and perform  the integration with logarithmic accuracy. Note that the lower
limit of integration is changed, 
$r_{\omega} \to r_{\omega}'=
r_{\omega}\left[1+2(Zt_{\perp}/\epsilon_0)^2
\ln\left(2\epsilon_0/\omega\right)\right]$.
After simple calculations we obtain
\begin{eqnarray}
\label{rq}
 \frac{\sigma_{H > H_f}}{\sigma_{H < H_f}} & = &   1 + \left(\frac{Zt_{\perp}}{\epsilon_0}\right)^{2}
 \ln{\left(\frac{2\epsilon_0}{\omega}\right)} 
\ \frac{10 + 8 V_{\omega} }{1+V_{\omega} } , \\
&&\kappa r_{\omega}  \ll  \epsilon_0/(2Zt_{\perp}). \nonumber
\end{eqnarray}
From now on we use the value $Zt_{\perp}/\epsilon_0=0.06$, which fits well the dc MR
of La$_{2-x}$Sr$_x$CuO$_4$  \cite{kotov06a}, and 
 we take $\epsilon_0 = 10 \mbox{meV}$.
  
\begin{figure}[t]
\centering
\includegraphics[height=185pt, keepaspectratio=true]{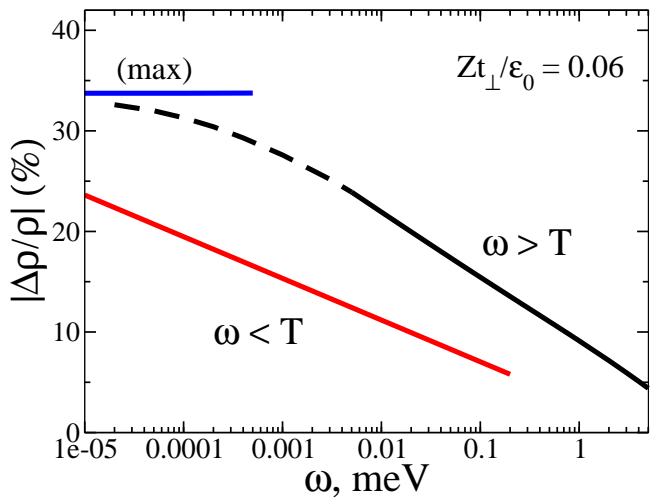}
\caption{(Color online.) Relative resistivity jump at  the spin
flop transition versus frequency. In the quantum limit $\omega > T$, 
the black solid line corresponds to the Intermediate large distance regime
described by Eq.~(\ref{rq}), and is justified at $\omega \gtrsim 10^{-2}\mbox{meV}$.
The horizontal blue line, labeled ``max", corresponds to the Very large distance regime
(ultra-low frequency) described by Eq.~(\ref{vld}).
The smooth crossover between these two limiting cases is shown 
schematically by the dashed line.
 At high temperatures $\omega < T$, the relaxation absorption formula
  Eq.~(\ref{rst1}) is plotted in red within its region of applicability;
it is expected to saturate at $\omega \sim 10^{-8}\mbox{meV}$ (not shown).}
\label{Fig1}
\end{figure}

As the frequency decreases, the ratio $(\sigma_{H > H_f}/\sigma_{H < H_f})$ 
slowly increases, following Eq.~(\ref{rq}).
 This behavior is illustrated in Figure \ref{Fig1},
where we have plotted the magnitude of the MR jump, i.e.
the quantity  $|\Delta \rho/\rho \equiv (\rho_{H > H_f}/\rho_{H < H_f}) -1|$. 
\noindent
The Intermediate large distance formula (\ref{rq}) fails somewhere below
$\omega \sim 10^{-2} \mbox{meV}$.
At lower frequencies one needs to perform the integration
in Eq.~(\ref{se}) using  the exact numerical expression for $P(r)$ derived in
Ref.~\cite{kotov06a}. This is expected to provide a  smooth crossover to the ultra-low
frequency asymptotic value given  by Eq.~(\ref{vld}) and shown in 
Figure~\ref{Fig1} by the horizontal blue line.

{\it{Jump in the in-plane ac conductivity at the spin-flop transition for 
field $H \parallel \hat{c}$. Phonon-dominated regime, $\omega < T$.}}
The regime $\omega > T$ considered in the previous section is not very
realistic, since experimentally it is easier to achieve $\omega < T$. To calculate the
 ac conductivity in this limit we use  the Austin-Mott
approach based on the relaxation mechanism \cite{AM}.
On the technical side we follow Ref.~\cite{Efros}.
In this regime there is a local quasistatic equilibrium established in an 
external ac electric field between two hole bound states separated by distance
$r$. The relaxation comes from the phonon assisted tunneling, and the 
characteristic relaxation time $\tau$ is
\begin{equation}
\frac{1}{\tau(r)} = \nu P(r),
\end{equation}
where $\nu\sim 1-10 \mbox{meV}$ is a characteristic phonon frequency, and $P(r)$ is the tunneling
probability considered previously. 
According to Eq.(4.7) of  Ref.~\cite{Efros} the conductivity is, adapted
 to our 2D case
\begin{equation}
\sigma(\omega) \propto \omega^{2} \int_{0}^{\infty} \frac{r^{3} 
\tau dr}{1+\omega^{2}\tau^{2}} \ G(r,T),
\end{equation}
where the function $G(r,T)$ is expected to be independent of  $r$ at large
 distances, and $G(r,T) \propto T$. We have then
\begin{equation}
\label{ps}
\sigma(\omega)  \propto (\omega^{2}/\nu) \ T  \int_{0}^{\infty} \frac{r^{3} 
P(r)dr}{P^{2}(r)+(\omega/\nu)^{2}}.
\end{equation}
Define 
\begin{equation}
\label{ro1}
\tilde{r}_{\omega} = \frac{1}{2\kappa}\ln{(\nu/\omega)}.
\end{equation}
Similarly to the previous section we have the
Very large distance regime, $\kappa \tilde{r}_{\omega} \gg \epsilon_0/(2Zt_{\perp})$, and 
the Intermediate large distance regime, 
$\epsilon_0/(2Zt_{\perp}) \gg\kappa  \tilde{r}_{\omega} \gg 1$.
The typical hopping distance in this case,  given by (\ref{ro1}), 
 contains an extra  factor 1/2, compared to (\ref{roo}). 
We take $\nu = 5 {\mbox{meV}}$. Therefore 
the Very large distance regime  sets in round $10^{-8} {\mbox{meV}}$,
 i.e. in the KHz range.
The  Intermediate large distance regime on the other hand
is realized at $10^{-5} {\mbox{meV}}$ (i.e. MHz range) and higher.

Using (\ref{pid}) and performing the integration in Eq. (\ref{ps}), we 
obtain the conductivity after the spin flop in the Intermediate large 
distances regime
\begin{equation}
\label{c1}
\sigma(\omega)  \propto \omega \ T \left (\frac{1}{2\kappa} \right ) \tilde{r}_{\omega}^{3} 
\left [ 1+ 10 \left(\frac{Zt_{\perp}}{\epsilon_0}\right)^{2} (\kappa \tilde{r}_{\omega}) 
\right ] \ .
\end{equation}
Therefore the conductivity ratio is, for  $\kappa \tilde{r}_{\omega} \ll \epsilon_0/(2Zt_{\perp})$
\begin{equation}
\label{rst1}
 \frac{\sigma_{H > H_f}}{\sigma_{H < H_f}} =  1 +5 
 \left(\frac{Zt_{\perp}}{\epsilon_0}\right) ^{2} \ln{(\nu/\omega)} \ .
\end{equation}
To make the consideration complete we also present the  result for 
 the Very large distance regime, $\kappa \tilde{r}_{\omega} \gg \epsilon_0/(2Zt_{\perp})$,
\begin{equation}
\label{rst}
\frac{\sigma_{H > H_f}}{\sigma_{H < H_f}}=\left(1-\frac{4Zt_{\perp}}{\epsilon_0}\right)^{-2} \ .
\end{equation}
Our discussion so far does not account for the Coulomb interaction.
Presumably for strong interaction $e^{2}/\tilde{r}_{\omega} \gg T$,
easily achievable for not too small $\omega$ and not too high $T$, the
conductivity (\ref{c1}) crosses over to a temperature-independent
behavior, as  discussed in Ref.~\cite{Efros}. 
However, the form of  Eq.~(\ref{rst1}) remains valid, where the coefficient
5 is replaced by 4. Within the accuracy of our calculations
(in particular the value of $\nu$) the difference can  be ignored. 
We also mention that the exponent in
 Eq.~(\ref{rst}) changes from -2 to -3/2 in the strong interaction case
(i.e. the result becomes  the same as  Eq.~(\ref{vld}) for $V_{\omega} \gg 1$).

We are only aware of old work \cite{Chen} that has studied the problem
 experimentally  in La$_2$CuO$_{4+y}$
(and not theoretically explained until now) in the
 high-temperature regime, $\omega \ll T$;
our results in the relaxation regime are completely consistent
 with those measurements.
The weak ferromagnetism of La$_2$CuO$_{4}$ provides a unique opportunity to
study the interplay between the charge and  spin dynamics.
The jump of the in-plane ac conductivity at the spin flop transition
is the most direct manifestation of this interplay.
Therefore further experimental studies of the jump could provide  important confirmation
of our  emerging understanding of lightly doped insulating La$_{2-x}$Sr$_x$CuO$_4$.

In conclusion, we have developed a description of the ac  magnetotransport 
across a spin flop  transition in the strongly localized regime
(low $T$ and $\omega$ compared to the hole's binding energy $\sim 10{\mbox{meV}}$).
 Figure~\ref{Fig1}, which roughly spans the MHz to THz frequency range,
summarizes the typical main features of our results, namely:
(1.) Slow, logarithmic variation over a wide  frequency
range (Eqs.~(\ref{rq},\ref{rst1})), (2.)
 Existence of an upper limit for the MR at low  $\omega$ (Eq.~(\ref{vld})),
(3.) Different magnitudes of the MR and saturation frequencies, depending on the nature
 of the ac VRH  mechanism (resonant absorption or relaxational).
We hope that the rich ac behavior found in this work also stimulates
additional experiments, as both our theoretical understanding 
 and sample quality have improved dramatically during the last few years.
 Magnetotransport in the strongly localized regime is a case
 where the presence of disorder actually simplifies considerably
 the problem of spin-charge interplay, and consequently our theory
 applies mostly to the La family of cuprates.

We are grateful to A.H. Castro Neto and D.K. Campbell for stimulating discussions.
 V.N.K. was supported by Boston University. O.P.S. gratefully  acknowledges
support from the Alexander von Humboldt Foundation.

\end{document}